\documentclass[12pt]{iopart}
\usepackage{epsfig}
\begin{document}
\title{Transport properties of graphene nanoribbons with side-attached organic molecules}
\author{L. Rosales$^1$, M. Pacheco$^1$\cite{email}, Z. Barticevic$^1$, A. Latg\'e$^2$,
and P. A. Orellana$^3$}
\address{$^1$Departamento de F\'isica, Universidad Santa Mar\'ia,
Casilla 110 V, Valpara\'iso, Chile}
\address{$^2$Instituto de F\'isica, Universidade Federal Fluminense, 24210-340,
Niter\'oi-RJ, Brazil}
\address{$^3$Departamento de F\'isica, Universidad Cat\'olica del Norte, Casilla 1280,
Antofagasta, Chile}
\date{\today}

\begin{abstract}

In this work we address the effects on the conductance of graphene
nanoribbons (GNRs) of organic molecules adsorbed at the ribbon edge.
We studied the case of armchair and zigzag GNRs with quasi
one-dimensional side-attached molecules, such as linear
poly-aromatic hydrocarbon and poly(para-phenylene). These
nanostructures are described using a single-band tight-binding
Hamiltonian and their electronic conductance and density of states
are calculated within the Green's function formalism based on
real-space renormalization techniques. We found that the conductance
exhibits an even-odd parity effect as a function of the length of
the attached molecules. Furthermore, the corresponding energy
spectrum of the molecules can be obtained as a series of Fano
antiresonances in the conductance of the system. The latter result
suggests that GNRs can be used as a spectrograph-sensor device.


\end{abstract}
\maketitle

\section{Introduction}

The high electronic mobility found in graphene nanoribbons (GNRs)
and the facilities for their growth suggest they may be used for
future electronics and  in  many others  nanotechnological
applications. GNRs are single atomic layers which can be understood
as an infinite unrolled carbon nanotube\cite{Nakada,pimenta,novo1}.
The special electronic behavior of GNRs, defined by their quasi
one-dimensional electronic confinement and the shape of the ribbon
ends, indicates remarkable applications in graphene-based
devices\cite{Shi}. GNRs share many of the electronic, mechanical,
and thermal properties of carbon nanotubes\cite{castro}, however,
due to the flat structure they seem to be easier to manipulate than
carbon nanotubes\cite{PJH}.

Similar to the case of nanotubes, ballistic transport and quantized
electronic conductance are expected to be found in graphene
structures. In particular, different quantization rules have been
predicted for clean ZGNRs and AGNRs\cite{Peres}. Edge states present
in zigzag ribbons provide a single channel for electron conduction
which is not the case for armchair configuration\cite{wakabayashi}.

Distinct and amusing designs of GNRs are proposed to highlights
their peculiar transport properties. One example is the
manifestation of the so-called Klein paradox\cite{Shi}, which
predicts that the electron can pass through a high potential barrier
without an exponential decay. Also interesting is the discussion
about the possible manifestation of half-metallicity  in
nanometer-scale GNRs reported on recent first-principles
calculations\cite{Son}. From the point of view of applications, it
has been shown that graphene exhibits chemical sensors properties.
Actually, it was recently reported a graphene-based gas sensor,
which allows the detection of individual gas molecules adsorbed on
graphene\cite{novo2,berger}.

\begin{figure}[h!]
\begin{center}
\includegraphics[width=8.5cm,height=8cm]{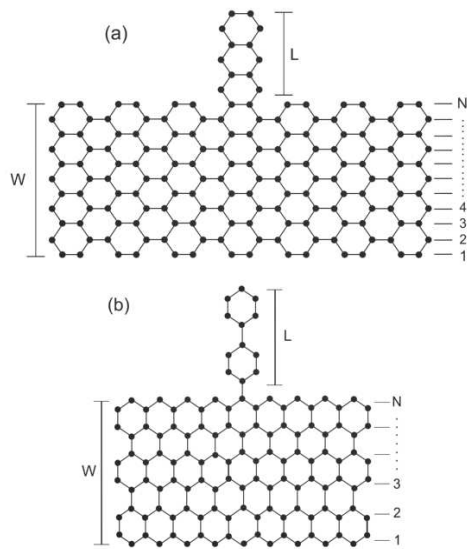}
\end{center}
\caption{Schematic view of an (a) AGNR and a (b) ZGNR of width $W$
with a molecule of length $L$ attached at the edges of the ribbons.}
\end{figure}

In this work, we will focus on the effects of side-attached
one-dimensional chains of hexagons pinned at the edges of the GNRs.
These one dimensional chains could be useful to simulate,
qualitatively, the effects on the electronic transport of GNRs when
benzene-based organic molecules are attached into the edges of the
ribbons. A simple scheme is proposed to reveal the main electronic
properties and the changes on the conductance of such decorated
planar structures. For simplicity, we consider armchair (AGNRs) and
zigzag (ZGNRs) nanoribbons and  linear poly-aromatic hydrocarbon
(LPHC) and poly(para-phenylene) as the organic molecules.

\section{Model}

Schematic diagrams of the two types of GNRs treated here are shown
in Fig.1. The attached molecules are simulated by simple
one-dimensional carbon hexagonal structures connected to the GNRs.
Following a tight-binding approach we adopt a single $\pi $-band
Hamiltonian, taking into account only nearest-neighbors hopping
interaction,
\begin{equation}
H_{T}=\sum_{i}\varepsilon _{i}c_{i}^{\dag }c_{i}+\gamma
\sum_{\left\langle ij\right\rangle }\left( c_{i}^{\dag
}c_{j}+h.c\right)\,\,,
\end{equation}
where $\varepsilon _{i}$ is the on-site energy (taken as the zero of
energies) and $\gamma=2.75 eV$ is the hopping parameter. Although
being a simple model, it was chosen to highlight the general trends
of the transport features in GNRS due to edge perturbations. To
describe an actual benzene-based molecular structures, one should
simulate single and double bonds, which may be done by changing
hopping parameters\cite{Chen} in an alternate sequence.

The local density of states (LDOS) and the conductance of the
systems are obtained within the Green's function formalism based on
real-space renormalization techniques\cite{rocha}. The conductance
is calculated via the Landauer formula and following a standard
surface Green's functions matching scheme\cite{Nardelli}. In this
scheme LDOS at site $i$ can be written as
\begin{equation}
\rho _{i}\left( \omega \right) =-\frac{1}{\pi } Im\left[
G_{ii}\left( \omega \right) \right]\,\,.
\end{equation}
Within the linear response approach the conductance can be written
in term of the system Green's function\cite{Garcia}, as
\begin{equation}
G=\frac{2e^{2}}{h}T\left(E_{F}\right)=\frac{2e^{2}}{h}Tr\left[
\Gamma_{L}G_{C}^{R}\Gamma_{R}G_{C}^{A}\right]\,\,,
\end{equation}
where $G_{c}^{R,(A)}$ is the retarded (advanced) conductor Green's
function, $T\left(E_{F}\right)$ is the transmission function of an
electron crossing through the central conductor, $\Gamma
_{L,R}=i\left[ \Sigma ^{L,R}-\left( \Sigma ^{L,R}\right) ^{\dagger
}\right]$ is the coupling between the leads and the conductor and
$\Sigma ^{L,R}$ are the self-energies of each lead.

\section{Results and discussion}

Results for the conductance as a function of the Fermi energy for an
(a) 11-AGNR and a (b) 6-ZGNR are shown in Fig.2. The curves from
down to up are for different lengths $L$ of the attached molecules.
The typical conductance steps for pristine GNRs are now marked by
the presence of well defined suppressions and/or dips at particular
energy values, which evolve depending on the length $L$ of the
linear molecules. These dips are known as  Fano antiresonances
(FARs), and emerge in a system when discrete states coexist with
continuum energy states\cite{fano1,fano2}.

\begin{figure}[h!]
\begin{center}
\includegraphics[width=10.0cm,height=7cm]{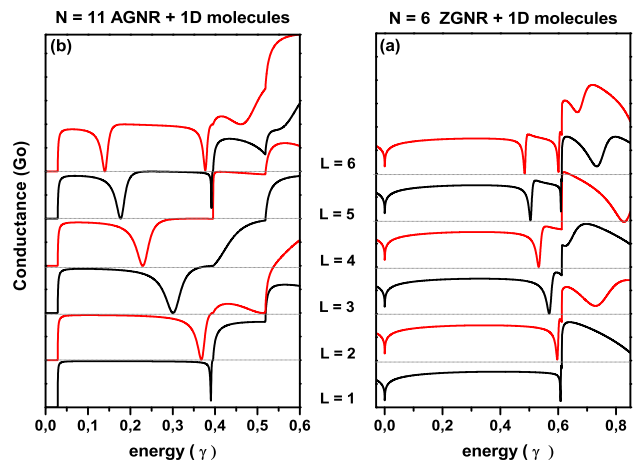}
\end{center}
\caption{Conductance as a function of the Fermi energy for an (a)
11-AGNR and a (b) 6-ZGNR  with  one-dimensional molecules attached
to the ribbon, as shown in Fig.1. The curves from bottom to top
correspond to increasing  molecular lengths L. For a better
visualization, all curves are displaced by a quantum of conductance
Go.}
\end{figure}

In order to highlight the main effects that the presence of the
organic molecules impress over the ribbons, we calculate the LDOS
and the conductance of the smallest systems which actually may be
viewed as a graphene stripe. We have chosen a $N=2$ ZGNR and a $N=5$
AGNR (this last one is not the narrowest ribbon but it is one with
the smallest gap). Results for both considered cases are shown in
Fig.3, where we have changed the length of the organic molecules
from $L=1$ to $L=6$ (upper panel 5-AGNR and down panel 2-ZGNR). We
note that, even for the smallest molecule considered,  the typical
Fermi level divergence (localized state) in the LDOS of the pristine
ZGNR (shown in dashed lines in Fig. 3(c)), completely disappears as
a result of the perturbation.  We also observe that for all molecule
lengths considered, the conductance of the system exhibits complete
suppressions for particular values of the Fermi energy. These
energies, as we will show below, coincide with the spectrum of the
corresponding attached molecule.

We note that the FARs of the conductance are smooth and wide except
when the Fermi energy coincides with one of the Van Hove singularity
in the LDOS in which case the FAR are very sharp. For those energies
an even-odd parity effect can be observed in the conductance as a
function of the molecule length. This effect arises from the
intrinsic properties of the one-dimensionality of the systems, and
it is in straight analogy with odd-even parity effects manifested in
atomic wires\cite{rubio,fclaro,orellana,kim}. In our particular
narrow ribbons this effect occurs at energies around $0,8\gamma$ for
the AGNR  and $1,0\gamma$ for ZGNR. For wider ribbons the effect
should be noticeable for lower energies.

\begin{figure}[h!]
\begin{center}
\includegraphics[width=10.0cm,height=6.5cm]{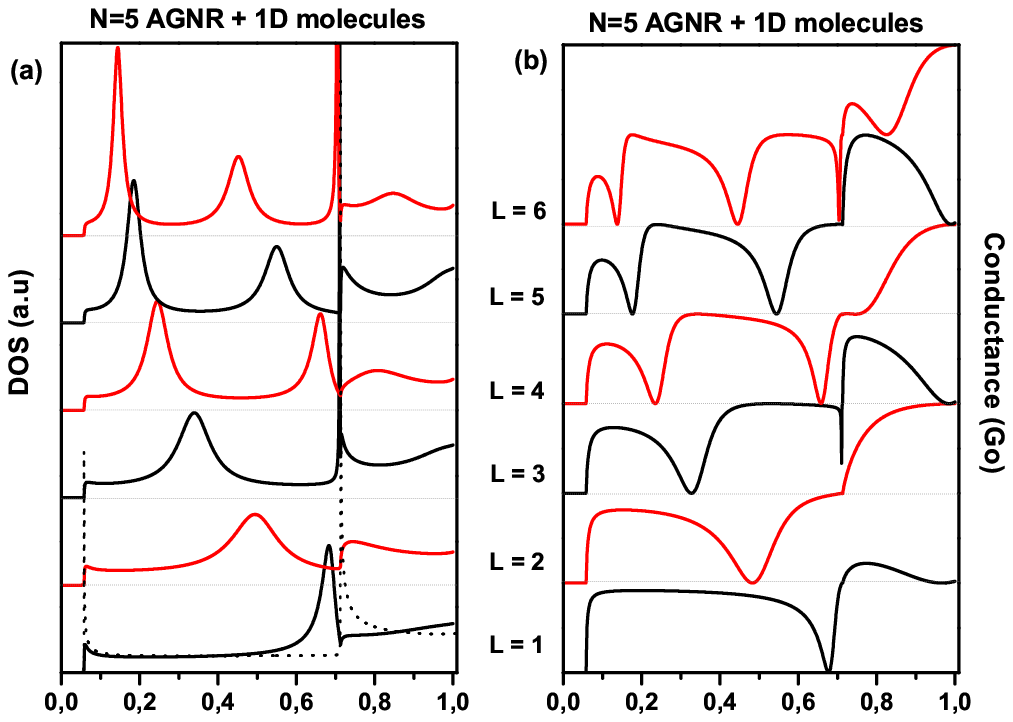}
\includegraphics[width=10.0cm,height=6.5cm]{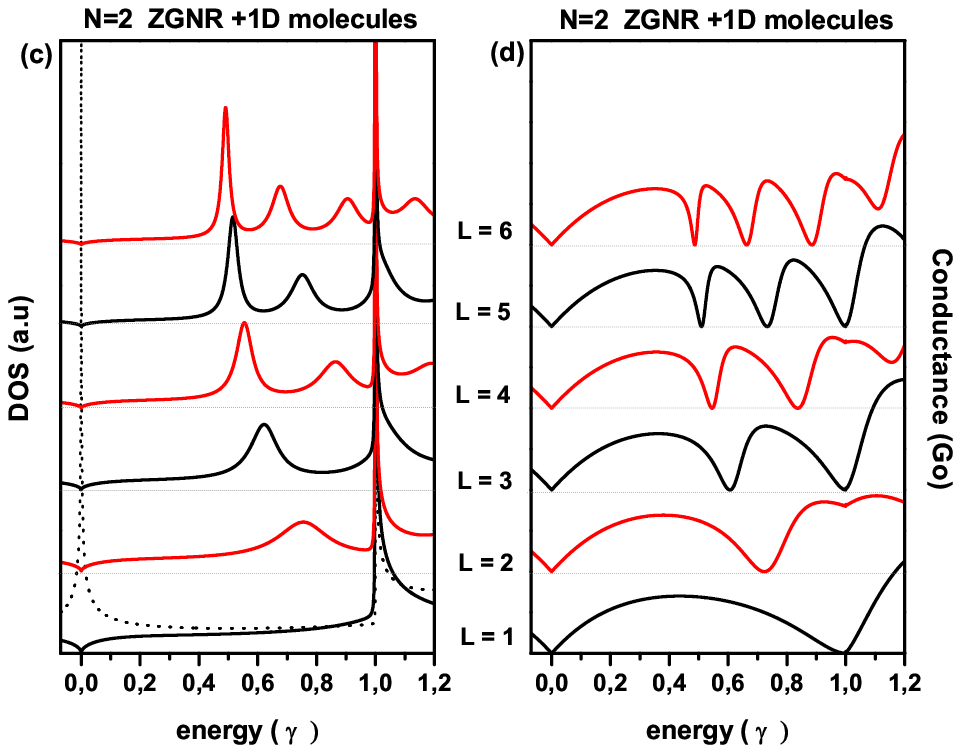}
\end{center}
\caption{LDOS and Conductance for an  5-AGNR (upper panels) and a
2-ZGNR (down panels), with L increasing from 1 to 6 as marked in the
figures. All curves are displaced for a better visualization.}
\end{figure}

To analyze the origin of the FARs in the ribbons conductance spectra
when quasi one-dimensional molecules are attached to their edges, we
compare the behavior of the energy spectra of the isolated molecules
with the FAR energy positions. Results of both systems are displayed
in Fig.4 (upper and down panel), taking into account ribbons with
side-attached organic molecules of different lengths ($L=1$ up to
$L=11$). Except for the antiresonance at the Fermi level due to the
interface state of the ZGNRs, or the small gap in the case of the
AGNR, the conductance dips reveal exactly the normal modes of the
specific attached molecules.  This effect arises from the
hybridization of the quasi-bound levels of the organic molecules and
the coupling to the ribbons. A similar effect was found by Orellana
\emph{et al}.\cite{orellana} in quantum wires with side attached
nanowires.

\begin{figure}[h!]
\begin{center}
\includegraphics[width=10cm,height=6.0cm]{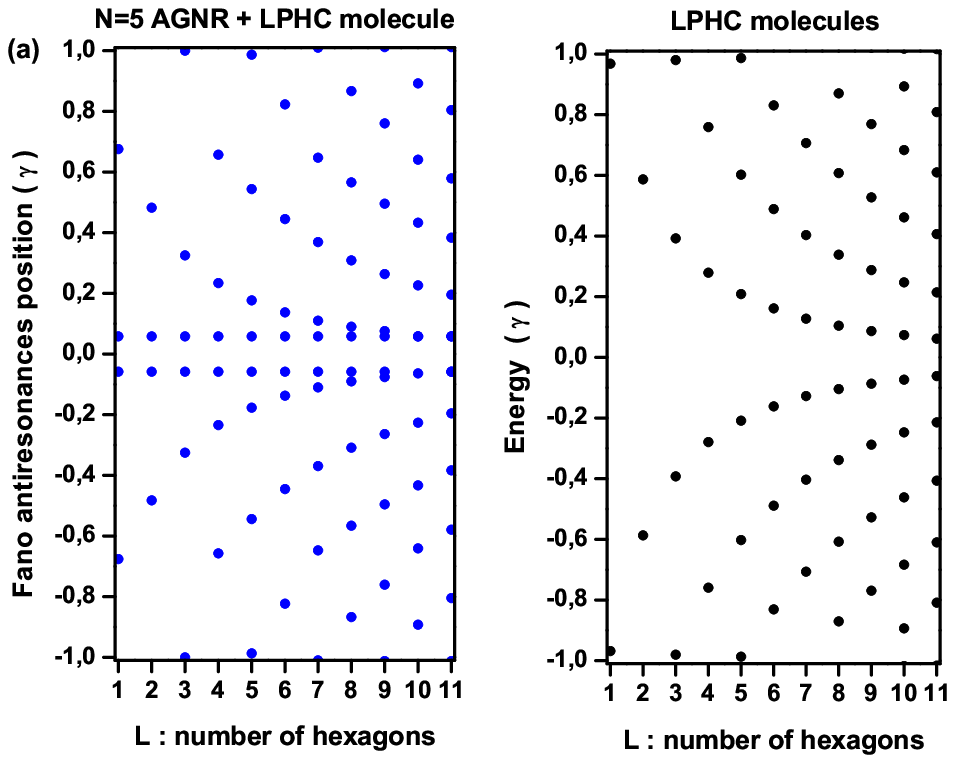}
\includegraphics[width=10cm,height=6.0cm]{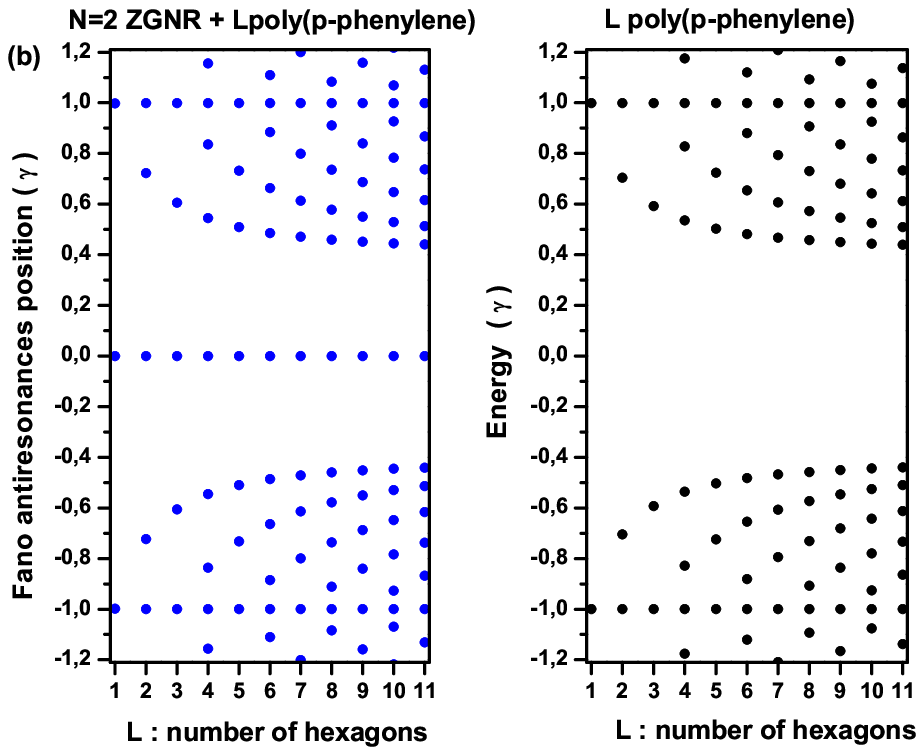}
\end{center}
\caption{Comparative plots between the position in energy of Fano
antiresonances in the conductance, and the corresponding energy
spectra of the isolated molecules, as a function of the molecule
length given in hexagons units.(a) 5-AGNR + LHPC molecules, (b)
2-ZGNR + poly(para-phenylene).}
\end{figure}

To investigate the robustness of the effect, we have calculated the
position of the conductance dips for ribbons with different widths
and we have compared it with the corresponding energy states of an
attached molecule of a given length. We show in Fig.5 results of the
conductance dips as a function of an  aspect ratio, defined as
$(W/L)$, for an N-AGNR with a side attached octacene $L=8$ molecule
in (a) and an N-ZGNR with an octa(para-phenylene) molecule in (b).
The position of the first energy states of the corresponding
molecule is included with a dashed line. It is clear that for the
cases of ribbons with zigzag edges the first energy levels of the
attached molecule perfectly coincide with the FAR position in the
conductance, for quite large aspect ratios. The same behavior has
also been observed for other molecule lengths. In the case of
ribbons with armchair edges we observe that the position of the FARs
exhibits fluctuations around the value of the molecule energy level,
as a function of the aspect ratio.
\begin{figure}[h!]
\begin{center}
\includegraphics[width=10.0cm,height=6.0cm]{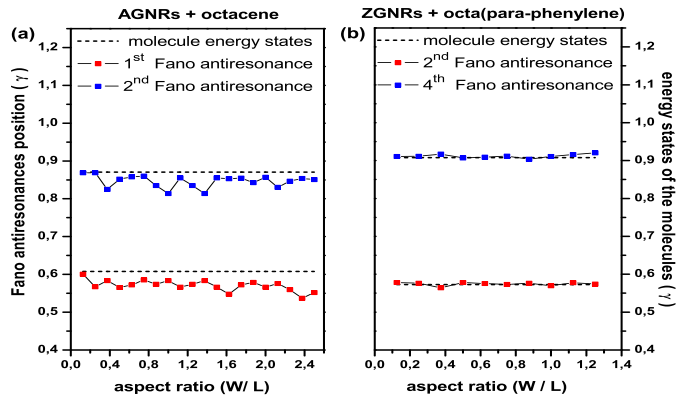}
\end{center}
\caption{Position in energy of the conductance Fano-antiresonances,
as a function of the aspect ratio (W/L), for an (a) AGNR+octacene
and (b) ZGNR+octa(para-phenylene), respectively. The lowest energy
states of the isolated molecule are shown with a dashed line.}
\end{figure}
This last feature is a consequence of the different electronic
behaviour of both type of ribbons, and most important, due to
differences in which the molecules are pinned to the ribbon edges.
We see in Fig.1 that for a ZGNR the attached molecule owns just a
single contact point with the ribbon edge. In the case of a molecule
attached to an AGNRs, the structure is more complex, with two
contact points, allowing more than a single way for hybridization of
the electronic states. On the other hand, ZGNR are always metallic
and the discrete energy states of the molecule may always interfere
with the quasi-continuum spectrum of the ZGNR. Contrarily, AGNRs are
always semiconductors, with an energy gap defined by the ribbon
width $W$\cite{Son2}, therefore holding fewer available states for
interference effects.

\section{Summary}
In summary, we studied the effects on the conductance of GNRs with
side-attached chains of hexagonal molecules. These topological
structures seem to be useful to describe, qualitatively, the effects
on the transport properties of GNRs when benzene-based organic
molecules of different length are attached to the ribbon edges. The
conductance of the GNRs reflects the energy spectrum of the quasi
one-dimensional system, suggesting that GNRs can be used as an
spectrograph-sensor device. Additionally, an even-odd parity effect
as a function of the length of the attached molecules, can be
observed in the conductance of these systems. Based on these first
results, one may propose an extended and more detailed study of
these nanostructures, using more sophisticated theoretical pictures.
An interesting task would be to investigate the transport of a large
number of molecules randomly distributed along the ribbon edges, and
also the inclusion of the charge distribution effects due to those
perturbations. A systematic analysis following this line may be
useful to determine the type and concentration of foreign entities
which could be detected with these kind of structures.

We acknowledge the financial support of CONICYT/Programa
Bicentenario de Ciencia y Tecnologia (CENAVA, grant ACT27),
brazilian agencies CNPq, FAPERJ, Rede Nacional de Nanotubos, grant
of Instituto do Mil\^enio. A.L. and L.R. thank the hospitality of
UTFSM-Valpara\'{\i}so and UFF-Brazil, respectively, where part of
this work was done.

\end{document}